\documentclass[aps,prl,preprint,superscriptaddress]{revtex4}

\usepackage{graphicx}% Include figure files
\usepackage{dcolumn}% Align table columns on decimal point
\usepackage{bm}% bold math

\begin{document}

%\preprint{}

\title{Telecom-Band Entanglement Generation for Chipscale Quantum Processing}



%\affiliation{}
\author{Kim Fook Lee, and Prem Kumar}
\email[]{kflee@ece.northwestern.edu}
%\thanks{ }
\affiliation{Center for Photonic Communication and Computing, EECS
Department, \\ Northwestern University, 2145 Sheridan Road,
Evanston, IL, 60208, USA}
\author{Jay E. Sharping, Mark A. Foster,
and Alexander L. Gaeta}
\affiliation{School of Electrical and
Computer Engineering, Cornell University, Ithaca, NY 14853}
\author{Amy C. Turner, and Michal Lipson}
\affiliation{School of Applied and Engineering Physics, Cornell
University, Ithaca, NY 14853}

% Collaboration name if desired (requires use of superscriptaddress
% option in \documentclass). \noaffiliation is required (may also be
% used with the \author command).

\date{October 29, 2006}

\begin{abstract}
We demonstrate polarization-entanglement for non-degenerate and
degenerate photon-pairs generated through Kerr-nonlinearity in a
nano-scale silicon-on-insulator (SOI)waveguide. We use a compact
counter propagating configuration to create two-photon
polarization-entangled state, $|H\rangle |H\rangle + |V\rangle
|V\rangle$. We observe two-photon interference with visibility $>
91\%$ and $>80\%$ for non-degenerate and degenerate photon-pairs,
respectively. The experimental structure can be implemented on
optical chips as an integrated source of entangled photons for
future quantum computer and communication applications.
\end{abstract}

%\keywords{}

\maketitle

%\section{}
%\subsection{}
%\subsubsection{}

Four-photon scattering (FPS) in a dispersion shifted fiber has
proven to be an essential nonlinear process for providing both
non-degenerate and degenerate telecom-band photon-pairs in many
proof of principle
experiments~\cite{Sharping01,Fiorentino02,XLi05,XLi04,kflee06,jchen06,liang06}.
Briefly, two pump photons at frequency $\omega_p$ scatter through
the four-photon scattering to create signal($\omega_s$) and idler
($\omega_i$) photons, such that $2\omega _p=\omega _s+\omega _i$ and
$2\,k _p=\,k _s+\,k_i$. This process is always referred to
non-degenerate FPS. While in the reverse degenerate FPS, one pump
photon at signal frequency ($\omega_s$) and one pump photon at idler
frequency ($\omega_i$) will annihilate together to create two
degenerate photon-pair at frequency ($\omega_p$). For practical
quantum communication, non-degenerate entangled photon-pairs are
desirable for entanglement distribution in a wavelength-division
multiplexing (WDM) environment~\cite{OFC05,Li04b}, and enabling
parallel multi-users access. While the degenerate entangled
photon-pairs are suitable for quantum computers because quantum
logic gates are essentially complicated interferometers which
usually require quantum interference of two indistinguishable
photons such as Hong-Ou-Mandel interference~\cite{Ou87}.

The advanced of nano-technology in fabrication~\cite{Book03} and
atto-second light source in light manipulation~\cite{Tex05} motivate
the possibility of implementing practical quantum information and
schemes in an optical chip. Silicon photonics is a most recent
promising technology for providing a integrated optics
platform~\cite{Mihal05}. Waveguide confinement in SOI waveguides can
yield net anomalous group-velocity dispersion and enable four-wave
mixing process in classical domain~\cite{Gaeta06}. Prior to this
present work, we have observed the generation of correlated
photon-pairs through the FPS in the SOI waveguide~\cite{Jay06}.
However, nonclassical correlation and polarization-entanglement of
the generated photon-pairs have not been investigated.

\begin{figure}
\includegraphics[width=3.25in]{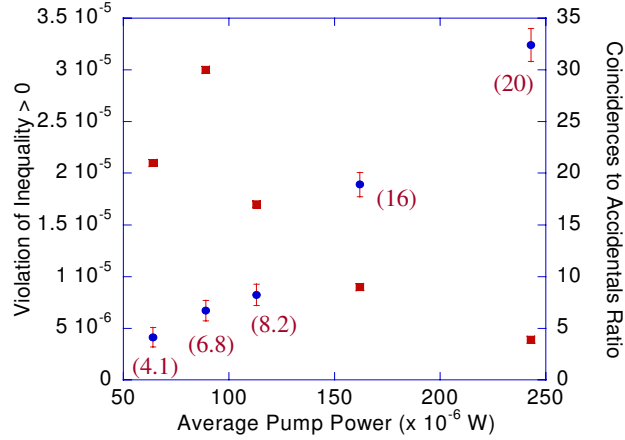}
\caption{\label{f1}Measurement on the violation of classical
inequality in Eq.~\ref{eq1} (circle) and coincidences to accidentals
ratio (square) for the generated signal-idler photon pairs in the
SOI waveguide. The number in the parenthesis indicates the order of
violation over standard deviations of measurement uncertainty.}
\end{figure}

In this work, we extend our study to demonstrate nonclassical nature
of two-photon coincidences of the generated signal-idler
photon-pairs in the SOI waveguide. For this purpose, we use the
similar experiment setup as shown in ref[~\cite{Jay06}]. Briefly, we
launch a linearly polarized pump light into SOI waveguide, then the
generated signal-idler photon-pair are spectrally filtered out
through their individual WDM filters and hence provides signal and
idler quantum channels. In order to prove that the photon-pairs
exhibit nonclassical correlation, we need to violate classical
inequality which is valid for two classical light sources, as given
by ~\cite{Zou91},
\begin{eqnarray}
R^{s,i}_{c}-R^{s,i}_{ac}-2(R^{s/2}_{c}-R^{s/2}_{ac}+R^{i/2}_{c}-R^{i/2}_{ac})\leq
0,\label{eq1}
\end{eqnarray}
where $R^{s,i}_{c}$ and $R^{s,i}_{ac}$ are the coincidence and
calculated accidental coincidence count rates for the signal and
idler channels. The lights in signal and idler channels are assumed
to be the two classical light sources. $R^{s/2}_{c}$ and
$R^{s/2}_{ac}$ ($R^{i/2}_{c}$ and $R^{i/2}_{ac}$) are the
coincidence and calculated accidental coincidence count rates
measured by passing the signal (idler) channel through a 50/50
beamspliter, respectively. For nonclassically correlated
signal-idler photon pairs, the term $R^{s,i}_{c}-R^{s,i}_{ac}$ in
Eq.~\ref{eq1} is the true two-photon coincidences which will be
proportional to photon-pair production rate. While the term
$(R^{s/2}_{c}-R^{s/2}_{ac}+R^{i/2}_{c}-R^{i/2}_{ac})$ in
Eq.~\ref{eq1} vanishes because the probability for observing a
photon-pair in either signal or idler channels is zero. We measure
the inequality as a function of average pump power in the waveguide
as shown in Fig.~\ref{f1}. We observe larger violation of the
inequality as we increase the average pump power which corresponds
to the increase of photon-pair production rate. Our experimental
result confirms that the signal-idler photon-pair generated through
FPS in the waveguide is indeed nonclassically correlated. From the
same experimental data, we plot coincidences to accidentals ratio
(CAR) and obtain the peak ratio of $\simeq 30$. At the peak ratio
where the average pump power is about $90\mu W$ , the Eq.~\ref{eq1}
yields $(67\pm 10)\times10^{-7}$ which provides the violation over 6
standard deviations. One should adopt CAR as the best estimate for
the purity of the correlated photon-pairs before the source can be
used for many practical quantum processing, where the
accidental-coincidence counts contributed from background photons
are inevitable.

In the second experiment, we study the generation of
polarization-entangled photon-pairs through non-degenerate and
degenerate FPS in a SOI waveguide. Since the FPS process only occurs
in the transverse electric (TE) mode of SOI waveguide, we need to
modify the counter-propagating scheme
(CPS)~\cite{Kumar05,XLi06,Takesue04} so that the created two-photon
polarization-entangled state propagates in the direction opposite to
the incoming input pump field. We observe two-photon interference
with visibility $>91\%$ and $\>80\%$ for non-degenerate and
degenerate cases, respectively. These results are promising for
realization of chip, and future quantum computer and communication
applications.

\begin{figure}
\includegraphics[width=3.25in]{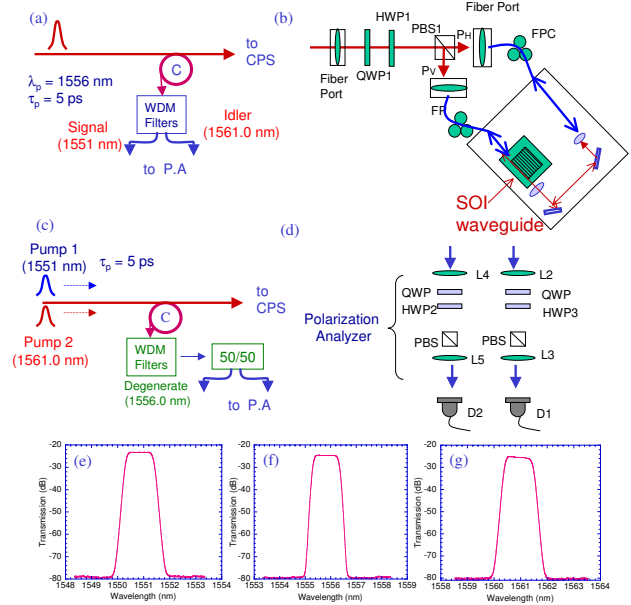}
\caption{\label{f2}A schematic of the experimental setup. (a) The
input pump field for nondegenerate FPS. (b). The CPS with SOI
waveguide. (c). The two pump fields for degenerate FPS. (d).
Polarization analyzers for signal and idler photons. FP, fiber-port;
LP, linear polarizer; L2, L3, L4, L5, fiber-to-free space
collimators; PBS, polarization beamsplitter; HWP, QWP, half- and
quarter-wave plates; FPC, fiber polarization controller. The WDM
filter shape for (e) the signal, (f) the pump and (g) the idler.}
\end{figure}

As shown in Fig.~\ref{f2}(a), the pump is a mode-locked pulse train
with pulse duration $\simeq\,5$\,ps and repetition rate of
50.3\,MHz. The pump is initially obtained from a femtosecond fiber
laser by using a WDM filter with 1\,nm full-width at half-maximum
(FWHM) passband. The pump central wavelength is at
$1555.9\,\textrm{nm}$. To achieve the required power, the pump
pulses are further amplified by an erbium-doped fiber amplifier
(EDFA). The amplified spontaneous emission (ASE) from the EDFA is
suppressed by passing the pump through the two cascaded WDM filters.
The pump power goes through a fiber circulator(C)) to the CPS. The
configuration setup for the SOI waveguide is shown as a small box in
Fig.~\ref{f2}(b). The waveguide is a 9.33-mm long and its cross
section is 600 nm x 300 nm. It has built-in an inverse taper at both
ends. We couple the light into the waveguide using a lensed fiber.
The polarization of the input field has to be aligned parallel to
the TE mode of the waveguide. We then use a 3 mm focal-length
aspheric lens to collect the output light from the waveguide. A
regular free-space to fiber collimator is used to couple the light
back into fiber. Before the CPS is used to create two-photon
polarization-entangled state, we want to make sure all birefringence
sensitive devices including SOI waveguide and polarization analyzers
are properly aligned and compensated. Briefly, we use half
wave-plate (HWP1) and quarter wave-plate (QWP1)in the CPS to adjust
the input pump field so that it is vertically polarized at the PBS1.
This pump will travel in the counter-clockwise direction of the CPS.
Fiber polarization controller (FPC1) is used to transform the pump
field into the TE mode of the waveguide. Then,  the polarization of
the generated signal and idler photons together with the pump photon
in the counter-clockwise direction in the CPS are transformed by the
FPC2 to horizontal polarization. They transmit through the PBS1 in
the direction opposite to the incoming pump field and back to the
circulator. The signal and idler photons are filtered out and sent
to the polarization analyzers (PA). A set of QWP and HWP in the PA
is used to compensate the polarization birefringence effects
experienced by the signal and idler photons since they traveled from
the PBS1, so that they are retained to horizontal polarization
before the PBS. The angle settings of HWP2 and HWP3 in signal and
idler channels are then referred to the detection polarization
angles $\theta_1=0$ and $\theta_2=0$. Now, if the HWP1 is rotated to
$45^\circ$, the input pump field is transformed to horizontally
polarized field. This pump field propagates in the clockwise
direction of the CPS. With the same polarization transformation
provided by the FPC2 and FPC1, the generated signal and idler
photons in the clockwise direction are transformed to vertical
polarization and reflected at the PBS1 to the opposite direction of
incoming pump field. In the backward direction, one should realize
that the HWP1 at $45^\circ$ will again transform the vertical
polarization state of the reflected lights to horizontal
polarization state. The independent optimizations of the electric
field polarization alignments in counter-clockwise and clockwise
directions of this CPS and SOI waveguide are essential due to the
fact that the FPS only occurs in TE mode.

We now rotate the HWP1 to $22.5^\circ$ so that the input pump field
at the PBS1 is now split into two equally powered, orthogonally
polarized components $P_H$ and $P_V$. For low four-photon scattering
efficiencies, where the probability for each pump pulse to scatter
more than one pair is low, the clockwise and counter-clockwise pump
pulses scatter signal/idler photon-pairs with probability amplitudes
$|H_i\rangle |H_s\rangle$ and $|V_i\rangle|V_s\rangle$,
respectively. After propagating through the SOI waveguide, these two
amplitudes of the photon-pair are then coherently superimposed
through the same PBS1. This common-path polarization interferometer
has good stability for keeping zero relative phase between
horizontally and vertically polarized pumps, and hence is capable of
creating polarization-entanglement of the form $|H_i\rangle
|H_s\rangle + |V_i\rangle |V_s\rangle$ at the input port of the
PBS1. Since the two-photon polarization-entangled state propagates
in the opposite direction of input pump field, one should realize
that the HWP1 at $22.5^\circ$ again rotates the two-photon state to
$45^\circ$ basis. This backward action can be compensated at the
signal and idler channels by adding the initial angle settings of
HWP2 and HWP3 by the amount of $22.5^\circ$. Note that one could
keep the initial angle settings of HWP2 and HWP3 and account the
backward action of HWP1 as projecting the signal and idler photons
to $45^\circ$ prior to coincidence detection. Since the correlation
function of the entangled state $|H_i\rangle |H_s\rangle +
|V_i\rangle |V_s\rangle$ is $\cos^{2}(\theta_1-\theta_2)$, either
option as mentioned above gives maximum coincidence counts. To
reliably detect the scattered photon-pairs, an isolation between the
pump and signal/idler photons in excess of 100\,dB is required. We
achieve this by using two cascaded WDM filters with FWHM of about
1\,nm in the signal and idler channels, which provide total pump
isolation greater than 110\,dB. The selected signal and idler
wavelengths are $1550.95\,\textrm{nm}$ and $1561.0\,\textrm{nm}$,
respectively, corresponding to $\simeq\, 5.0\,\textrm{nm}$ detuning
from the pump's central wavelength. The photon-counting modules
consist of InGaAs/InP avalanche photodiodes operated at a rate of
780\,kHz, which is downcounted by $1/64$ from the original pump
pulses. The total detection efficiencies for the signal and idler
photons are about $0.7\%$ and $0.8\%$, respectively.

\begin{figure}
\includegraphics[width=3.25in]{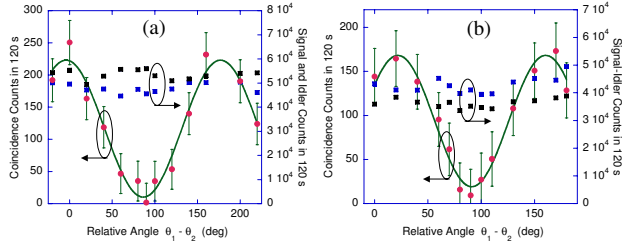}
\caption{\label{f3} The measured two-photon interference for (a)
non-degenerate and (b) degenerate polarization entangled
photon-pairs with visibility $> 91\%$ and $> 80\%$, respectively.}
\end{figure}

We measure two-photon interference to justify the purity of the
entangled photon-pairs generated by SOI waveguide. The average power
for each pump in the waveguide is $96\mu W$ after including the
$80\%$ coupling efficiency of the fiber-port, $80\%$ transmission of
the FPC1 and also $10\%$ coupling efficiency from the tapered fiber
to waveguide. We fix $\theta_1$, and vary $\theta_2$, and record
single counts for both signal and idler channels as well as
coincidence counts between the two channels for each value of
$\theta_2$. We observe TPI with visibility $> 91\%$ as shown in
Fig.~\ref{f3}(a), without subtraction of accidental-coincidences.
The fitting function used in the figure is
$\cos^{2}(\theta_1-\theta_2)$. The photon-pair production rate is
about 0.08. The visibility of TPI is in agreement with our current
and previous measurements~\cite{Jay06} on CAR of 25-30 for the
correlated non-degenerate photon-pairs generated in the same
waveguide.

We further study the purity of degenerate entangled photon-pair with
the CPS and SOI waveguide. For this purpose, we use the similar
experimental scheme~\cite{jchen06} to prepare two input pump fields.
Briefly, we spectrally filter out the desired pump central
wavelengths at $\omega_{pl}$ =1550.95 nm and $\omega_{p2}$ =1560.01
nm by using a double-grating filter (DGF) with FWHM of 0.8 nm. These
two pumps are combined together and then amplified by using EDFA.
After the EDFA, the two input pump fields are further filtered out
by a second DGF. We then combine them in a WDM by using the
following configuration as ($P+R \rightarrow C$). We also make sure
the two independent input pump fields to be overlapped in time and
equal power, so that the phase matching condition of the reverse
degenerate FPS is optimized in the waveguide. The degenerate
two-photon polarization-entangled state is created by the similar
manner in the CPS as described in the non-degenerate case. The
degenerate photon-pairs at wavelength of 1555.9 nm is filtered out
by using two cascaded WDM filters. In order to bring the degenerate
photon-pairs into two spatially separated detectors, we need to use
a 50/50 fiber beamsplitter after the WDM filters as shown in
Fig.~\ref{f2}(c). By doing this, we post-select the case where the
photon-pair splits leading to coincidence detection. In other words,
after the 50/50 beamsplitter, we use coincidence measurement to
post-project the state $|H\rangle |H\rangle + |V\rangle |V\rangle$.
The average power for each pump used in this scheme is $288\mu W$
which corresponds to photon-pair production rate of 0.12. We measure
two-photon interference and observe TPI with visibility $> 80\%$ as
shown in Fig.~\ref{f3}(b).

All the above measurements are made without subtracting the
accidental coincidence counts. Even though literature~\cite{Qlin06}
predicts that spontaneous Raman scattering process (RS) in the TE
mode of the waveguide could exhibit a narrow peak with a FHWM of
0.8nm located at about 100nm away from the pump, but so far no
photon-counting measurement is conducted to measure how the RS
spreads from the narrow peak to the pump. We believe that the Raman
scattering photons prevent this entangled source to achieve TPI with
unit visibility. Since it is a chipscale waveguide, it can be easily
cooled to suppress RS photons. And also, the detection efficiencies
in the experiments are low due to the coupling loss, which could be
one of the main reason in degradation of the two-photon
entanglement. Degenerate photon-pairs exhibit lower TPI visibility
compared with non-degenerate photon-pairs because we use higher
average power for two pump fields through the EDFA which could
create in-band ASE noise photons to the detection.

In conclusion, we have demonstrated the generation of
polarization-entanglement for non-degenerate and degenerate
photon-pairs in a SOI waveguide. We believe the nano-scale silicon
based entanglement source will lead to realistic implementations of
quantum communication and computing protocols. This work is
supported in part by the NSF under Grant No.\ EMT- 0523975.

\begin{acknowledgments}
This work is supported in part by the NSF under Grant No.\ EMT-
0523975.
\end{acknowledgments}

%\bibliographystyle{apsrev}
%\bibliography{thesis}
% pasted in .bbl file dsf-pol-ent-pk.bbl

\newcommand{\noopsort}[1]{} \newcommand{\printfirst}[2]{#1}
  \newcommand{\singleletter}[1]{#1} \newcommand{\switchargs}[2]{#2#1}

\end{document}